\documentclass[%
reprint,
 amsmath,amssymb,
 aps,
 english
]{revtex4-1}

\usepackage{setspace}
\usepackage[utf8]{inputenc} 
\usepackage[english]{babel} 
\usepackage{graphicx}
\usepackage{latexsym}
\usepackage{color}
\usepackage[table]{xcolor}
\usepackage{amsmath,amsthm,amssymb}
\usepackage{accents}
\usepackage{flafter}
\usepackage{amssymb}
\usepackage{bbold} 
\usepackage{bm}
\usepackage{pstricks}
\usepackage{pst-plot}
\usepackage{cancel}
\usepackage{eufrak,mathrsfs,mathrsfs}
\usepackage{dsfont}    
\usepackage{makeidx}         
\usepackage{nomencl}         
\usepackage{float}
\usepackage{xfrac}
\usepackage{hyperref}
\usepackage{xcolor}
\usepackage{animate}

\definecolor{red}{rgb}{1.0,0.0,0.0}
\definecolor{blue}{rgb}{0.0,0.0,1.0}
\definecolor{dark-gree}{rgb}{0.0,0.5,0.0}
\hypersetup{
    colorlinks, 
    linkcolor={red},
    citecolor={blue}, 
    urlcolor={dark-gree}
}

\usepackage{graphicx}
\usepackage{dcolumn}

\newcommand{\eq}[1]{\begin{equation}#1\end{equation}}
\newcommand{\tn}[1]{\textnormal{#1}}
\newcommand{\lb}[1]{\label{#1}}

\newcommand{\ket}{\rangle}
\newcommand{\dst}{\displaystyle}
\newcommand{\x}{\hat{x}}
\newcommand{\y}{\hat{y}}
\newcommand{\z}{\hat{z}}

\begin{document}


\title{Comparing the Schr\"odinger and Dirac Descriptions of an Electron in a Uniform Magnetic Field}

\author{ David Velasco Villamizar}
\email{david.velasco.v@gmail.com}
\affiliation{Departamento de F\'isica, Universidade Federal de Santa Catarina, Santa Catarina, CEP 88040-900, Brazil}%

\author{Benjamin Russell}
\email{br6@princeton.edu}
\affiliation{Frick Laboratory, Princeton University, Princeton NJ 08544, US}%




\date{\today}

\begin{abstract}
In this article we present a detailed description of an electron in a uniform magnetic field evolving under the Schr\"odinger equation using ladder operators.
Based on this analysis, we describe the same physical system using the Dirac equation, known from relativistic quantum mechanics.
The main differences between these two quantum mechanical approaches are discussed and we observe specifically how the relativistic phenomena modify the description of this particular quantum system.
\end{abstract}

\maketitle

\section{\label{sec:introduction}Introduction}

In this article we present the detailed calculations required for a didactic comparison between the dynamics of the Schr\"odinger and Dirac equations for an electron in a uniform magnetic field.
This is done in order to juxtapose these two quantum mechanical descriptions of the same system as a pedagogic tool as this comparison is not found in any quantum mechanics textbook.
For that reason, this analysis would be an invaluable resource to instructors, who could make use this example, either for an advanced undergraduate quantum mechanics course or for a beginning graduate course.

It is well known that one of the simplest phenomenon at the quantum scale is the interaction between an electron (which is a spin one half fermion) and an external uniform axial magnetic field \cite{Peter_Rowe,Tannoudji, Haugset, Yoshioka, Thaller, Greenshields}.
Although this physical system comprises only a single particle, whether it is at rest or in motion, its description is valuable as it furnishes the basic concepts required to understand other phenomena.
For specific examples see the electron vortex beams \cite{Babiker,Chowdhury,Bliokh} and the interaction of solid state materials with magnetism, known as the integer Quantum Hall effect \cite{Landau,Feldman_Kahn,Pekka,laughlin,Schattschneider}.
Additionally, this physical system shows an isomorphism to quantum optics, demonstrating a narrow relation to describe the Gaussian beam profile of electromagnetic radiation, leading to orthogonal states known as Orbital Angular Momentum of light \cite{Ole}.
This optical phenomenon has been obtained experimentally, due to the high accuracy now achievable and its great applicability \cite{Allen} of simulations in innumerable quantum systems of interest, for example, the orbital angular momentum Hall effect\cite{Zhang}, which is an analogue to the integer quantum Hall effect.

A brief outline of the paper is as follows, in section \ref{sec:II} we describe the interaction between spin $1/2$ and a uniform magnetic field.
In section \ref{sec:III} we explain in detail the Schr\"odinger equation of the electron in motion under the interaction of the external magnetic field, performing the analysis by using the ladder operator method.
Consequently, we can notice that the wave function of the electron exhibit a cylindrical symmetry, this feature is related with a uniform external magnetic field.
Also, we explain briefly the equivalence between the electron eigenstates and the Laguerre-Gaussian modes, which are characteristic of the orbital angular momentum of light in cylindrical coordinates.
We then further present a short discussion about the electron energy levels and the Landau levels.
Finally, in section \ref{sec:IV} we present a relativistic analysis using the Dirac equation for the electron wave function and discuss the main differences to the Schr\"odinger equation description.

\section{Spin-magnetic field interaction}\lb{sec:II}

Here we consider the quantum description of an electron initially at rest in a uniform external magnetic field.
We consider the electron magnetic dipole moment $\vec{\mu}\!=\!-\resizebox{0.075\hsize}{!}{$\left(\!\dst\frac{e}{m_{0}}\!\right)$}\vec{\hat{S}}$, where ($e\!=\!1,6021\!\!\times\!\!10^{-19}$C) is the modulus of the fundamental electric charge and the vectorial spin operator $\vec{\hat{S}}\!=\!\frac{\hbar}{2}\vec{\hat{\sigma}}$, where $\vec{\hat{\sigma}}\!=\!(\hat{\sigma}_{x},\hat{\sigma}_{y},\hat{\sigma}_{z})$ is the vector of Pauli matrices.
Neglecting any external force, the Hamiltonian of the electron is an interaction potential of the magnetic dipole moment with the magnetic field is $\hat{H} \!=\! -\vec{\mu}\!\cdot\!\vec{B}$.
To simplify the calculations we assume the orientation of the magnetic field is along the $\hat{z}$ axis, accordingly the electrons Hamiltonian is given by,
\eq{
 \hat{H} = \frac{e\hbar B}{2m_{0}}\left(\begin{array}{lr} 1&0\\0&-1\end{array}\right).
}
The energy levels are $E_{\pm} \!=\! \pm\hbar\omega/2$, where $\omega\!=\!\frac{eB}{m_{0}}$ is the precession frequency of the magnetic dipole momentum around the external field, known as Larmor Frequency.
In addition, the eigenstate of positive energy is $|0\ket\!\!=$\resizebox{0.06\hsize}{!}{$\left(\!\begin{array}{c} 1\\0 \end{array}\!\right)$} and the negative energy is $|1\ket\!\!=$\resizebox{0.06\hsize}{!}{$\left(\!\begin{array}{c} 0\\1 \end{array}\!\right)$}. These eigenstate are related with the spin orientation when it is aligned or anti-aligned to the magnetic field, respectively.

\section{Schr\"odinger equation of an electron in magnetic field} \lb{sec:III}
In contrast to last section, here we make a quantum description of an electron in motion, which is in a region with a uniform external magnetic field $\vec{B}$; in the classical description, the electron follows a helical path along the magnetic field showing an axial symmetry. According to this spatial feature, we apply a gauge transformation over the magnetic vector potential to take into account this symmetry. In particular, this specific symmetric gauge transformation is well know as the Landau gauge,
\eq{\lb{eq:12}
 \vec{A}(\vec{r}) = \frac{1}{2}\big( \vec{B}\times\vec{r} \big).
}
The linear momentum operator is $\hat{p}\!\rightarrow\!\hat{p}-q\vec{A}$, where the electron charge $q\!=\!-e$.
Then, in terms of momentum, the Hamiltonian is,  
\eq{\lb{eq:13}
\begin{split}
 \hat{H} &= \frac{1}{2m_{0}}\big( \hat{p} + e\vec{A} \big)^{2} + \frac{e}{m_{0}}\vec{B}\!\cdot\!\vec{S},\\
	&= \frac{-\hbar^{2}}{2m_{0}}\!\nabla^{2} \!+\! \frac{e^{2}A^{2}}{2m_{0}} \!-\! \frac{ie\hbar}{2m_{0}}\big( \nabla\!\cdot\!\vec{A} \!+\! \vec{A}\!\cdot\!\nabla \big) \!+\! \frac{e}{m_{0}}\vec{B}\!\cdot\!\vec{S}.
\end{split}
}
The second term of the Hamiltonian can be expanded,
\eq{\lb{eq:14}
 \begin{split}
 \frac{e^{2}A^{2}}{2m_{0}} &= \frac{e^{2}}{8m_{0}}\big( \vec{B}\times\vec{r} \big)^{2},\\
	&= \frac{e^{2}}{8m_{0}}B^{2}r^{2}\sin^{2}(\theta_{Br}),\\ 
	&= \frac{e^{2}}{8m_{0}} \big( B^{2}r^{2} - (\vec{B}\!\cdot\!\vec{r})^{2} \big),
 \end{split}
}
where $\theta_{Br}$ is the angle between the magnetic field and the electron vector position.
The third term in Eq. (\ref{eq:13}) will be calculated using,
\eq{\lb{eq:15}
 \nabla\!\cdot\!\big(\vec{A}\phi\big) = \big(\nabla\!\cdot\vec{A}\big)\phi + \vec{A}\!\cdot\!\nabla\phi.
}
The first term in last expression is equal to,
\eq{ \lb{eq:15_1}
 \begin{split}
  \big(\nabla\!\cdot\vec{A}\big)\phi &= \frac{1}{2}\nabla\!\cdot\! \big(\vec{B}\times\vec{r}\big)\phi,\\ 
	&= \frac{1}{2}\big(\nabla\times\vec{B}\big)\!\cdot\!\vec{r}\phi - \frac{1}{2}\big(\nabla\times\vec{r}\big)\!\cdot\!\vec{B}\phi,\\
	&=0.
 \end{split}
}
This result shows that the Landau gauge satisfies the Coulomb gauge.
Continuing the calculation of the second term in Eq. (\ref{eq:13}) in a similar way,
\eq{\lb{eq:16}
\begin{split}
 \vec{A}\!\cdot\nabla\phi &= \frac{1}{2}\big(\vec{B}\times\vec{r}\big)\!\cdot\!\nabla\phi,\\ 	
	&=\frac{1}{2} \vec{B}\!\cdot\!\big( \vec{r}\times\nabla \big)\phi. 
\end{split}
}
Representing the differential operator in term of the quantum linear momentum operator it follows that, 
\eq{\lb{eq:16a}
\begin{split}
 \vec{A}\!\cdot\nabla\phi &= \frac{i}{2\hbar} \vec{B}\!\cdot\!\big( \vec{r}\times\hat{P} \big)\phi,\\ 
	&= \frac{i}{2\hbar} \vec{B}\!\cdot\!\hat{L}\phi.
\end{split}
}
Eliminating $\phi$ from the above results allows us to write the Hamiltonian (\ref{eq:13}) as, 
\eq{\lb{eq:17}
 \hat{H}\!=\!\frac{-\hbar^{2}}{2m_{0}}\nabla^{2} \!+ \frac{e^{2}}{8m_{0}} \big( B^{2}r^{2} \!-\! (\vec{B}\cdot\vec{r})^{2} \big) + \frac{e}{2m_{0}} \vec{B}\!\cdot\! \big( \hat{L} + 2\hat{S} \big).
}
For simplicity and without loss of generality, we are going to assume a uniform magnetic field aligned along $\hat{z}$ axis.
Consequently, the Hamiltonian is can be written,
\eq{\lb{eq:18}
\begin{split}
 \hat{H} &= \frac{-\hbar^{2}}{2m_{0}}\nabla^{2} + \frac{e^{2}B^{2}}{8m_{0}} \big(x^{2}+y^{2}\big) + \frac{eB}{2m_{0}} \big( \hat{L}_{z} + 2\hat{S}_{z}\big),\\
	&= \frac{-\hbar^{2}}{2m_{0}}\frac{\partial^{2}}{\partial z^{2}} + \frac{-\hbar^{2}}{2m_{0}}\left(\frac{\partial^{2}}{\partial x^{2}}+\frac{\partial^{2}}{\partial y^{2}}\right) \\
	&\hspace{1.5cm}+ \frac{e^{2}B^{2}}{8m_{0}} \big(x^{2}+y^{2}\big) + \frac{eB}{2m_{0}} \big( \hat{L}_{z} + 2\hat{S}_{z}\big).
\end{split}
}
According to this result, we notice clearly the contribution of various terms of different natures in the Hamiltonian.
These are, a free particle along $\hat{z}$ axis, a two-dimensional harmonic oscillator in $x\!-\!y$ plane with its projection of angular momentum along $\z$ axis and, finally, the electron spin interaction with the magnetic field.
We can express $\psi(\vec{r},\sigma)$, the associated electron wave function, as the product of functions in according to their nature,
\eq{\lb{eq:20}
 \psi(\vec{r},\sigma) = F(x,y)e^{ip_{z}z/\hbar}\Gamma,
}
where the spinorial function is,
\eq{ \lb{eq:spinorial_function}
 \Gamma=\left\{\begin{array}{ll} {\tiny\left(\begin{array}{c} 1\\0\end{array}\right)},&m_{s}=+1/2\\ {\tiny\left(\begin{array}{c} 0\\1\end{array}\right)}, &m_{s}=-1/2\end{array}\right.,
} 
and $m_{s}$ is the quantum number equal to the spin projection along the magnetic field direction.
In virtue of this, we can analyze the two-dimensional harmonic movement of the electron, getting the Schr\"odinger equation of a quantum harmonic oscillator,
\eq{\lb{eq:21}
 \underbrace{\left[\frac{-\hbar^{2}}{2m_{0}} \! \left(\!\frac{\partial^{2}}{\partial x^{2}}\!+\!\frac{\partial^{2}}{\partial y^{2}}\!\right) \!+\! \frac{1}{2}m_{0}\omega^{2} \!\big(\!x^{2}\!+\!y^{2}\!\big)\!\right]}_{\dst H'_{xy}}\!F \!=\! E'F,  
}
where $\omega\!=\!\frac{eB}{2m_{0}}$ is the harmonic frequency.
The quantum harmonic oscillator is,
\eq{\lb{eq:22}
 \begin{split}
      E'&=\hbar\omega (n_{x}+n_{y}+1),\\ 
	&= \hbar\omega (n+1), \qquad \qquad n\!\geq\!0.
 \end{split}
}
We note that each energy level $E'_{n}$, is related to ($n+1$) degenerate eigenstates $F_{n_{x},n_{y}}$, 
\eq{\lb{eq:23}
 F_{n,0}\;,\; F_{n-1,1} \;,\;   \cdots\;,\;   F_{0,n}.
}
In relation to this degeneracy, we look for a physical quantity which enables a better characterization of states belonging to a certain energy level.
Given the spatial symmetry of the harmonic potential under a rotation around the $\z$ axis and the axial component of the orbital angular momentum of the electron $\hat{L}_{z} \!=\! \hat{x}\hat{p}_{y} \!-\! \hat{y}\hat{p}_{x}$, we can analyze the two-dimensional motion of the electron by applying the ladder operator method to each Cartesian coordinate,
\eq{\lb{eq:25}
 \begin{split}
	\hat{a}_{x}&=\frac{1}{\sqrt{2}}\left( \hat{x}\beta + i\frac{\hat{P}_{x}}{\beta\hbar} \right), \quad \hat{a}^{\dagger}_{x}=\frac{1}{\sqrt{2}}\left( \hat{x}\beta - i\frac{\hat{P}_{x}}{\beta\hbar} \right),\\
	\hat{a}_{y}&=\frac{1}{\sqrt{2}}\left( \hat{y}\beta + i\frac{\hat{P}_{y}}{\beta\hbar} \right), \quad \hat{a}^{\dagger}_{y}=\frac{1}{\sqrt{2}}\left( \hat{y}\beta - i\frac{\hat{P}_{y}}{\beta\hbar} \right),
 \end{split}
}
where the constant $\beta\!\!=\!\!\sqrt{\frac{m_{0}\omega}{\hbar}}$ is reciprocal to the natural length of the harmonic oscillator \cite{Tannoudji}.
Moreover, these operators satisfy the following commutation relations $[\hat{a}_{x},\hat{a}^{\dagger}_{x}]\!=\![\hat{a}_{y},\hat{a}^{\dagger}_{y}]\!=\!1$.
Furthermore, we can express $\hat{x}$, $\hat{y}$, $\hat{p}_{x}$, $\hat{p}_{y}$, as function of the ladder operators.
In that way, we have, 
\eq{\lb{eq:26}
\begin{split}
   \hat{x} &= \frac{1}{\beta\sqrt{2}}\big(\hat{a}_{x}+\hat{a}^{\dagger}_{x}\big), \qquad \hat{P}_{x} = -i\frac{\hbar\beta}{\sqrt{2}}\big(\hat{a}_{x}-\hat{a}^{\dagger}_{x}\big),\\
   \hat{y} &= \frac{1}{\beta\sqrt{2}}\big(\hat{a}_{y}+\hat{a}^{\dagger}_{y}\big), \qquad \hat{P}_{y} = -i\frac{\hbar\beta}{\sqrt{2}}\big(\hat{a}_{y}-\hat{a}^{\dagger}_{y}\big).
 \end{split}
}
The angular momentum operator is given by,
\eq{\lb{eq:27}
 \hat{L}_{z} = i\hbar \big( \hat{a}_{x}\hat{a}^{\dagger}_{y} - \hat{a}^{\dagger}_{x}\hat{a}_{y} \big).
}
Similarly, we can express the harmonic Hamiltonian as,  
\eq{\lb{eq:28}
 \hat{H}'_{xy} = \hbar\omega\big( \hat{a}^{\dagger}_{x}\hat{a}_{x} + \hat{a}^{\dagger}_{y}\hat{a}_{y} + 1 \big).
}
Calculating the commutation relation between the angular momentum and the Hamiltonian yields: 
\eq{\lb{eq:29}
 \begin{split}
   \big[\hat{H}'_{xy},\hat{L}_{z}\big] &=\! \bigg[ \hbar\omega\big(\!\hat{a}^{\dagger}_{x}\hat{a}_{x} \!+\! \hat{a}^{\dagger}_{y}\hat{a}_{y} \!+\! 1\!\big), i\hbar \big(\! \hat{a}_{x}\hat{a}^{\dagger}_{y} \!-\! \hat{a}^{\dagger}_{x}\hat{a}_{y} \!\big) \bigg],\\
	&=\! i\hbar^{2}\omega \big(\! -\!\hat{a}_{x}\hat{a}^{\dagger}_{y} \!+\! \hat{a}^{\dagger}_{x}\hat{a}_{y} \!+\! \hat{a}^{\dagger}_{y}\hat{a}_{x} \!-\! \hat{a}_{y}\hat{a}^{\dagger}_{x}),\\ 
	&=\! 0.
 \end{split}
}
Thus, we can show that $\hat{L}_{z}$ is a constant of motion.
Consequently, the angular momentum operator has the same set of eigenstates as $\hat{H}'_{xy}$; in this way, we can expose an axial symmetry in the system.
As such, we are going to implement the cylindrical coordinate system to describe the quantum operators corresponding to the electron position and momentum.
Additionally, we will borrow from geometrical optics the notion of the Jones' vectors for the circular polarization of light \cite{Jones_vector}.
One can introduce new ladder operators related to the axial symmetry $\hat{a}_{R}$ and $\hat{a}_{L}$, which are right and left circular operators respectively.
These are denoted as a function of the Cartesian ladder operators as,
\eq{\lb{eq:30}
 \hat{a}_{R} = \frac{1}{\sqrt{2}} \big( \hat{a}_{x} \!-\! i\hat{a}_{y} \big), \qquad \hat{a}_{L} = \frac{1}{\sqrt{2}} \big( \hat{a}_{x} \!+\! i\hat{a}_{y} \big).
}
Particularly, when these operators $\hat{a}_{R}$ and $\hat{a}_{L}$ act on the eigenstate $F_{n_{x},n_{y}}$, they generate a quantum state which is a linear combination of $F_{n_{x}-1,n_{y}}$ and $F_{n_{x},n_{y}-1}$.
This results in a state with energy smaller by $\hbar\omega$.
Contrastingly, $\hat{a}^{\dagger}_{R}$ and $\hat{a}^{\dagger}_{L}$ generate a quantum state with energy bigger by $\hbar\omega$.

Moreover, these operators satisfy the following commutation relations, 
\eq{\lb{eq:31}
 \big[\hat{a}_{R}, \hat{a}^{\dagger}_{R}\big] = \big[\hat{a}_{L}, \hat{a}^{\dagger}_{L}\big] = 1.
}
Hence, we can express,
\eq{\lb{eq:32}
 \begin{split}
	\hat{a}^{\dagger}_{R}\hat{a}_{R} = \frac{1}{2} \big( \hat{a}^{\dagger}_{x}\hat{a}_{x} + \hat{a}^{\dagger}_{y}\hat{a}_{y} + i(\hat{a}_{x}\hat{a}^{\dagger}_{y} - \hat{a}^{\dagger}_{x}\hat{a}_{y})\big),\\
	\hat{a}^{\dagger}_{L}\hat{a}_{L} = \frac{1}{2} \big( \hat{a}^{\dagger}_{x}\hat{a}_{x} + \hat{a}^{\dagger}_{y}\hat{a}_{y} - i(\hat{a}_{x}\hat{a}^{\dagger}_{y} - \hat{a}^{\dagger}_{x}\hat{a}_{y})\big).
 \end{split}
}
It leads us to express $\hat{H}'_{xy}$ and $\hat{L}_{z}$ as,
\eq{\lb{eq:33}
 \hat{H}'_{xy} = \hbar\omega\big( \hat{N}_{R} + \hat{N}_{L} + 1 \big), \qquad \hat{L}_{z} = \hbar\big( \hat{N}_{R} - \hat{N}_{L} \big), 
}
where $\hat{N}_{R}\!=\!\hat{a}^{\dagger}_{R}\hat{a}_{R}$ and $\hat{N}_{L}\!=\!\hat{a}^{\dagger}_{L}\hat{a}_{L}$ are known as the Hermitian number operator. When acting on the eigenstate $F_{n_{x},n_{y}}$, these operators simply return a positive integer $n_{R}$ and $n_{L}$, respectively. The physical meaning of these numbers are related to the current energy level, as we will show later.

\begin{figure}[t!]
   \centering
      \includegraphics[scale=0.40]{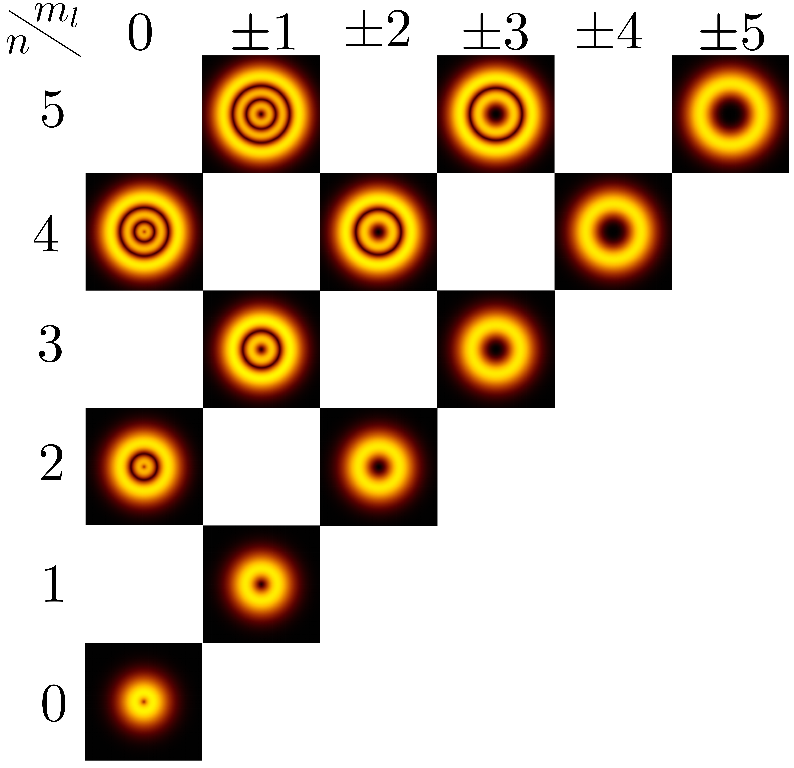}
   \caption{\small Electron radial probability density of eigenstates belonging of the first six values of $n$ and its individual set of $m_{l}$. All of these densities showing a null probability of finding the electron in the center position and some well defined rings, where the amount of these rings is $(n\!-\!|m_{l}|)/2$, which is equivalent to the number of roots of the generalized Laguerre polynomials.}
 \label{fig:Radial_distribution}
\end{figure}

If we denote the ground state as $F_{0,0}$, as the state of the lowest energy. In relation to this state, we can express any other eigenstate as a consecutive action of the operators $\hat{a}^{\dagger}_{R}$ and $\hat{a}^{\dagger}_{L}$. Thus, 
\eq{\lb{eq:34}
 \hspace{-0.3cm}F_{n_{R},n_{L}} \!=\! \dst\frac{1}{\sqrt{(n_{R})!(n_{L})!}} (\hat{a}^{\dagger}_{R})^{n_{R}}(\hat{a}^{\dagger}_{L})^{n_{L}} F_{0,0}.
}
Since $F_{n_{R},n_{L}}$ is a set of eigenstates in common to $\hat{H}'_{xy}$ and $\hat{L}_{z}$, we can obtain the energy levels $\hbar\omega(n\!+\!1)$ and the eigenvalue $m_{l}\hbar$.
In this way, we can define the main quantum number $n$ and the orbital quantum number $m_{l}$ as function of $n_{R}$ and $n_{L}$,
\eq{\lb{eq:35}
 n = n_{R} \!+\! n_{L}, \qquad  m_{l} = n_{R} \!-\! n_{L}.
}
Since $n_{R}$ and $n_{L}$ are positive integers, there are ($n\!+\!1$) degenerate eigenstate for every energy level, as shown in Eq. (\ref{eq:23}).
This proves the next relation, 
\eq{\lb{eq:36}
 \begin{array}{ll}
	n_{R} = n \; &; \; n_{L}=0\\
	n_{R} = n-1 \; &; \; n_{L}=1,\\
	\qquad\vdots & \\
	n_{R} = 0 \; &; \; n_{L}=n.\\
 \end{array}
}
Observing the particular case $n\!=\!0,1$ and its possible values of $m_{l}$, we have,
\eq{\lb{eq:37}
 \begin{split}
  n=0 &\quad\Rightarrow\quad \big\{ \;\;n_{R}=0 \;,\; n_{L}=0 \;\rightarrow\; m_{l}=0\\
  n=1 &\quad\Rightarrow\quad \left\{\begin{array}{l} n_{R}=1 \;,\; n_{L}=0 \;\rightarrow\; m_{l}=+1 \\ n_{R}=0 \;,\; n_{L}=1 \;\rightarrow\; m_{l}=-1 \end{array}\right. .
 \end{split}
}
In general, for every fixed value of $n$, there is a set of values of $m_{l}$,
\eq{\lb{eq:38}
 m_{l} = n \;,\; n-2 \;,\; n-4 \;,\; \cdots \;,\; -n+2 \;,\; -n.
}
In virtue of this transformation of notation, we can express the eigenstate $F_{n_{R},n_{L}}$ by using the pair of quantum numbers $n,m_{l}$, which are more appropriate to characterize this quantum system due to the observable $\hat{L}_{z}$. Thus, we have,
\eq{\lb{eq:39}
 F_{\resizebox{0.30\hsize}{!}{$\dst n_{R}\!=\!\frac{n\!+\!m_{l}}{2},n_{L}\!=\!\frac{n\!-\!m_{l}}{2}$}}.	
}
Then, we use the cylindrical coordinates,
\eq{\lb{eq:40}
 \begin{array}{l} 
   x = \varrho\cos\varphi \\ y = \varrho\sin\varphi \\ z=z
 \end{array}, \qquad \varrho\geq0,\quad 0\leq\varphi<2\pi,
}
We can rewrite the Hamiltonian of the two-dimensional harmonic oscillator in this coordinate system.
Also, we can transform the set of operator $\hat{a}_{R}$ and $\hat{a}_{L}$.
In that way, we obtain,
\eq{\lb{eq:41}
 \begin{split}
   \hat{a}_{R} &= \frac{1}{\sqrt{2}} \big( \hat{a}_{x} - i\hat{a}_{y}  \big),\\
	&=\frac{1}{2} \left[ \left(\hat{x}\beta + i\frac{P_{x}}{\beta\hbar}\right) - i\left(\hat{y}\beta + i\frac{P_{y}}{\beta\hbar}\right) \right],\\
	&=\frac{1}{2} \left[ \beta\big(\hat{x}-i\hat{y}\big) + \frac{i}{\hbar\beta}\big(\hat{P}_{x} - i\hat{P}_{y}\big) \right],\\
	&=\frac{1}{2} \left[ \beta\big(x-iy\big) + \frac{1}{\beta}\left(\frac{\partial}{\partial x} - i\frac{\partial}{\partial y}\right) \right].
\end{split}
}
The radial distance and the azimuthal angle are defined as $\varrho\!=\!\sqrt{x^{2}\!+\!y^{2}}$ and $\varphi\!=\!\tn{arctan}\left(y/x\right)$.
We can verify that,
\eq{\lb{eq:42}
  \big(x-iy\big) = \varrho\big(\cos\varphi -i\sin\varphi  \big) = \varrho e^{-i\varphi}.
}
and the Cartesian differentiation will be written as,
\eq{\lb{eq:43}
 \begin{split}
  \frac{\partial}{\partial x} &= \frac{\partial}{\partial\varrho}\frac{\partial\varrho}{\partial x} + \frac{\partial}{\partial\varphi}\frac{\partial\varphi}{\partial x} = \cos\varphi \frac{\partial}{\partial\varrho} - \frac{1}{\varrho}\sin\varphi \frac{\partial}{\partial\varphi},\\
  \frac{\partial}{\partial y} &= \frac{\partial}{\partial\varrho}\frac{\partial\varrho}{\partial y} + \frac{\partial}{\partial\varphi}\frac{\partial\varphi}{\partial y} = \sin\varphi \frac{\partial}{\partial\varrho} + \frac{1}{\varrho}\cos\varphi \frac{\partial}{\partial\varphi},
 \end{split}
}
Then, 
\eq{\lb{eq:44}
 \begin{split}
 \hspace{-2cm} \left(\!\frac{\partial}{\partial x} \!-\!i\frac{\partial}{\partial y}\!\right) \!&=\! \big(\!\cos\varphi \!-\!i\sin\varphi\!\big) \frac{\partial}{\partial \varrho}\\ 
	&\qquad-\frac{1}{\varrho}\big( \!\sin\varphi + i\cos\varphi \!\big)\frac{\partial}{\partial\varphi},\\ 
	&= e^{-i\varphi}\left(\! \frac{\partial}{\partial\varrho} - \frac{i}{\varrho} \frac{\partial}{\partial\varphi} \!\right).
 \end{split}
}
Thus, the operators $\hat{a}_{R}$ and $\hat{a}^{\dagger}_{R}$ are expressed in cylindrical coordinate as,
\eq{\lb{eq:45}
 \begin{split}
	&\hat{a}_{R} = \frac{e^{-i\varphi}}{2} \left[ \beta\varrho + \frac{1}{\beta}\frac{\partial}{\partial\varrho} - \frac{i}{\beta\varrho}\frac{\partial}{\partial\varphi} \right],\\ 
	&\hat{a}^{\dagger}_{R} = \frac{e^{i\varphi}}{2} \left[ \beta\varrho - \frac{1}{\beta}\frac{\partial}{\partial\varrho} - \frac{i}{\beta\varrho}\frac{\partial}{\partial\varphi} \right].
 \end{split}
}
Analogously, we can obtain,
\eq{\lb{eq:46}
 \begin{split}
	&\hat{a}_{L} = \frac{e^{i\varphi}}{2} \left[ \beta\varrho + \frac{1}{\beta}\frac{\partial}{\partial\varrho} + \frac{i}{\beta\varrho}\frac{\partial}{\partial\varphi} \right],\\
	&\hat{a}^{\dagger}_{L} = \frac{e^{-i\varphi}}{2} \left[ \beta\varrho - \frac{1}{\beta}\frac{\partial}{\partial\varrho} + \frac{i}{\beta\varrho}\frac{\partial}{\partial\varphi} \right].
 \end{split}
}
In particular, the action of $\hat{a}_{R}$ or $\hat{a}_{L}$ on the ground state $F_{n_{R}=0,n_{L}=0}$ is,
\eq{\lb{eq:47}
 \begin{split}
 \hat{a}_{R}F_{0,0} &= \frac{e^{-i\varphi}}{2} \left[ \beta\varrho + \frac{1}{\beta}\frac{\partial}{\partial\varrho} - \frac{i}{\beta\varrho}\frac{\partial}{\partial\varphi} \right]F_{0,0},\\ 
	&= 0.
 \end{split}
}
This result means physically the impossibility of annihilating a quantum of energy $\hbar\omega$ from the ground state. In virtue of this fact, we can solve the differential equation Eq. (\ref{eq:47}) and obtain the normalized eigenfunction,
\eq{\lb{eq:48}
 F_{0,0}(\varrho,\varphi) = \frac{\beta}{\sqrt{\pi}} e^{-\beta^{2}\varrho^{2}/2}.
}
In contrast, when $\hat{a}^{\dagger}_{R}$ acting $n_{R}$-times on $F_{0,0}(\varrho,\varphi)$ yields the creation of $n_{R}$ \emph{quanta} of energy $\hbar\omega$ on the ground state, 
\eq{\lb{eq:49}
 F_{n_{R},0}(\varrho,\varphi) = \frac{\beta}{\sqrt{\pi(n_{R})!}} \big(\beta\varrho\big)^{n_{R}} e^{-\beta^{2}\varrho^{2}/2}e^{in_{R}\varphi}.
} 
Similarly, the action of $\hat{a}^{\dagger}_{L}$ $n_{L}$-times on $F_{0,0}(\varrho,\varphi)$ yields, 
\eq{\lb{eq:50}
 F_{0,n_{L}}(\varrho,\varphi) = \frac{\beta}{\sqrt{\pi(n_{L})!}} \big(\beta\varrho\big)^{n_{L}} e^{-\beta^{2}\varrho^{2}/2}e^{-in_{R}\varphi}.
} 
After obtaining these results, it's worth mentioning that every energy level $E_{n}\!=\!\hbar\omega(n\!+\!1)$ corresponds to the maximum projection of the angular momentum $m_{l}\!=\!n$ in Eq. (\ref{eq:49}), or the minimum value $m_{l}\!=\!-n$  in Eq. (\ref{eq:50}).
Nonetheless, by a consecutive action of the operator $\hat{a}^{\dagger}_{L}$ $n_{L}$-times on the eigenfunction $F_{n_{R},0}$ or $\hat{a}^{\dagger}_{R}$ $n_{R}$-times on the eigenfunction $F_{0,n_ {L}}$, we obtain the eigenfunction $F_{n_{R},n_{L}}(\varrho,\varphi)$ for any pair of quantum numbers $n_{R}$, $n_{L}$.
Naturally, after calculating a lot of eigenfunctions by the action of the two creation operators, we start to obtain eigenfunctions which are proportional to the generalized Laguerre polynomials, denoted by $L^{(\alpha)}_{n}(x)$, multiplied by a Gaussian function which limits spatially the probability density of the electron in transversal direction\cite{Tannoudji,Yoshioka}.
In general, any wave function associated with the eigenstate ($n\!=\!n_{R}\!+\!n_{L}$) with angular momentum ($m_{l}\!=\!n_{R}\!-\!n_{L}$) can be written as follows,
\eq{\lb{eq:51}
 F_{n,m_{l}}(\varrho,\varphi) \!=\! C_{n}\beta \big(\beta\varrho\big)^{|m_{l}|} {\large L}^{|m_{l}|}_{\left(\frac{n-|m_{l}|}{2}\right)}\!(\beta^{2}\varrho^{2}) e^{-\beta^{2}\varrho^{2}/2} e^{im_{l}\varphi},
}
\begin{figure}[t!]
 \centering
  \includegraphics[scale=0.15]{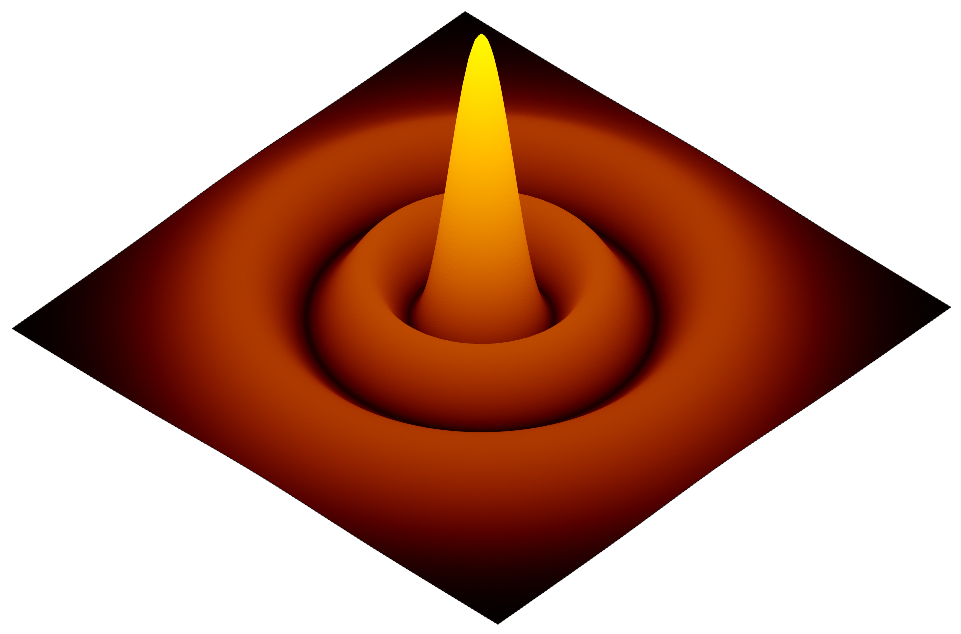}
 \caption{\small Electron spatial probability density for the eigenstate $F_{4,0}$. The two darkest rings are associated with forbidden radial positions of the electron. }
 \lb{fig:densidade_4-0}
\end{figure}
\\
where $\beta^{2}\varrho^{2}$ is the argument of the polynomial function.
The normalization constant is,
\eq{
 C_{n} = \frac{\dst(-1)^{\left(\frac{n-|m_{l}|}{2}\right)}\left(\dst\frac{n-|m_{l}|}{2}\right)!}{\sqrt{\pi\left(\dst\frac{n+|m_{l}|}{2}\right)!\left(\dst\frac{n-|m_{l}|}{2}\right)!}}.
}
\begin{table}[t!]
 \centering
\begin{tabular}{|c|c|l|} 
 \hline
 $n$ & $m_{l}$ & $F_{n,m_{l}}(\rho,\varphi)$\\ \hline
 0  & 0   & $\dst F_{0,0}=\frac{\beta}{\sqrt{\pi}}e^{-\beta^{2}\rho^{2}/2}$  \\ \hline
 1  & $1$   & $\dst F_{1,1}=\frac{\beta}{\sqrt{\pi}} (\beta\rho) e^{-\beta^{2}\rho^{2}/2} e^{i\varphi}$  \\ \hline
 2  & $2$   & $\dst F_{2,2}=\frac{\beta}{\sqrt{2\pi}} (\beta\rho)^{2} e^{-\beta^{2}\rho^{2}/2} e^{i2\varphi}$ \\ 
    & 0   & $\dst F_{2,0}=\frac{\beta}{\sqrt{\pi}} \big[(\beta\rho)^{2}-1\big] e^{-\beta^{2}\rho^{2}/2}$ \\ \hline
 3  & $3$   & $\dst F_{3,3}=\frac{\beta}{\sqrt{6\pi}}(\beta\rho)^{3}e^{-\beta^{2}\rho^{2}/2}e^{i3\varphi}$\\
    & $1$   & $\dst F_{3,1}=\frac{\beta}{\sqrt{2\pi}}\big[(\beta\rho)^{3}-2(\beta\rho)\big]e^{-\beta^{2}\rho^{2}/2}e^{i\varphi}$ \\ \hline
 4  & $4$   & $\dst F_{4,4}=\frac{\beta}{2\sqrt{6\pi}}(\beta\rho)^{4}e^{-\beta^{2}\rho^{2}/2}e^{i4\varphi}$\\
    & $2$   & $\dst F_{4,2}=\frac{\beta}{\sqrt{6\pi}}\big[(\beta\rho)^{4}-3(\beta\rho)^{2}\big]e^{-\beta^{2}\rho^{2}/2}e^{i2\varphi}$\\
    & 0   & $\dst F_{4,0}=\frac{\beta}{\sqrt{4\pi}}\big[(\beta\rho)^{4}-4(\beta\rho)^{2}+2\big]e^{-\beta^{2}\rho^{2}/2}$ \\ \hline
 5  & $5$   & $\dst F_{5,5}=\frac{\beta}{2\sqrt{30\pi}}(\beta\rho)^{5}e^{-\beta^{2}\rho^{2}/2}e^{i5\varphi}$\\
    & $3$   & $\dst F_{5,3}=\frac{\beta}{2\sqrt{6\pi}}\big[(\beta\rho)^{5}-4(\beta\rho)^{3}\big]e^{-\beta^{2}\rho^{2}/2}e^{i3\varphi}$\\
    & $1$   & $\dst F_{5,1}=\frac{\beta}{\sqrt{12\pi}}\big[(\beta\rho)^{5}-6(\beta\rho)^{3}+6(\beta\rho)\big]e^{-\beta^{2}\rho^{2}/2}e^{i\varphi}$ \\  \hline
 \end{tabular}
 \caption{Radial eigenfunction common to the two-dimensional harmonic oscillator and the observable $\hat{L}_{z}$, for the first six values of $n$. Negative angular momentum projection $m_{l}$ entails negative azimuthal phase.}
 \lb{tab:f_2D}
\end{table}
\\
The generalized Laguerre polynomials can be explicitly written using the formula,
\eq{\lb{eq:52}
 \hspace{-0.2cm} {\large L}^{|m_{l}|}_{\left(\!\frac{n-|m_{l}|}{2}\!\right)}\!\!(\beta^{2}\varrho^{2}) \!=\!\!\!\!\!\! \sum^{\left(\!\frac{n-|m_{l}|}{2}\!\right)}_{i=0} \!\!\!\!\!\! (-1)^{i} \!\! \left(\begin{array}{c} \!\!\! \left(\!\frac{n-|m_{l}|}{2}\!\right)\!+\!|m_{l}|\\ \left(\!\frac{n-|m_{l}|}{2}\!\right)\!-\!i\end{array} \!\! \right) \!\frac{\dst\big(\beta\varrho\big)^{2i}}{i!}.
}
Exemplifying a particular case in Figure \ref{fig:densidade_4-0}, we show the probability density of the eigenstate with null angular momentum $|F_{4,0}(\varrho,\varphi)|^{2}$.
Additionally, in Table \ref{tab:f_2D} we show the electron radial eigenfunction to the first six values of $n$.
In relation to these eigenfunctions, we obtain more information about every electron eigenstate when we calculate its radial probability density as,
\eq{
 D_{n,m_{l}}(\varrho) = 2\pi\varrho\big|F_{n,m_{l}}(\varrho,\varphi)\big|^{2}.
}
Thus, in Figure \ref{fig:Radial_distribution} we show the radial probability densities of every eigenstates belonging to the first six values of $n$.

We want to briefly mention a close relationship between the radial eigenfunction of the electron and Quantum Optics.
These eigenfunctions are the same as those we can obtain as a natural solutions of the wave equation under the paraxial approximation \cite{Goubau,Andrews}, for a Gaussian beam of light in cylindrical coordinates.
Commonly, in quantum optics these functions are called Laguerre-Gaussian modes, which are expressed through the generalized Laguerre polynomials $L^{|l|}_{p}$, where $p\!\ge\!0$ is the radial index which is related to the number of rings in the probability density of every eigenfunction; we can notice this number in Figure \ref{fig:Radial_distribution} along every diagonal from lower left to upper right.
The integer $l$ which is the azimuthal index, has the same physical meaning as the axial electron angular momentum. 

This physical feature lets us establish an isomorphic behavior between the quantum mechanics description of an electron in a uniform magnetic field and the Laguerre-Gaussian modes of light\cite{Ole}.
Where the amplitude of the paraxial beam of light corresponds to the spatial probability density of the electron and the Gouy phase of optics\cite{Simin_Feng} assumes the meaning of time for the quantum harmonic oscillator using an unitary dynamical evolution.
Recently, these Laguerre-Gaussian beams have been of considerable practical interest, particularly in the field of optical trapping, where they are applied to study the driving of micromachined elements with light as optical tweezers\cite{Paula}.

We can obtain the Schr\"odinger equation in cylindrical coordinates by using the full definition of the electron wave function,
\eq{\lb{eq:53}
 \begin{split}
	&\Big[ \hat{H}_{\rho\varphi}' \!+\! \frac{\hat{p}^{2}_{z}}{2m_{0}} + \hbar\omega\big(L_{z} \!+\! 2S_{z}\big) \Big]F_{n,m_{l}}e^{ip_{z}z/\hbar}\Gamma\\ 
	&\qquad\qquad= E \; F_{n,m_{l}}e^{ip_{z}z/\hbar}\Gamma.
 \end{split}
}
In this way, we obtain the total energy of the electron in relation to its eigenstates,
\eq{\lb{eq:54}
 E = \frac{p^{2}_{z}}{2m_{0}} + \hbar\omega\big( n + m_{l} + 2m_{s} + 1 \big),  \qquad \omega=\frac{eB}{2m_{0}}.
}
In general, the electron energy levels are $E\!\geq\!0$, being the zero of energy associated to eigenstates with a null linear momentum $p_{z}$ and the lowest possible projection of angular momentum $m_{l}\!=\!-n\!$ in relation to the main quantum number $n$, also, a spin orientation opposite to the magnetic field $m_{s}\!=\!-1/2$. Additionally, under some of the last conditions and getting free $m_{s}$, we can obtain the energy levels associated to the spin interaction with the external magnetic field as we shown in section \ref{sec:II}, but in this case we see the contribution of the zero-point energy which is related to the harmonic potential and that is the reason to have this energy shift.
\eq{\lb{eq:54_1}
 E = \frac{e\hbar B}{2m_{0}} \left( 2m_{s} + 1 \right).
}
Furthermore, we notice that the addition of $n$ and $m_{l}$ is always an even number.
Then, we can rewrite the energy expression using a new integer number,
\eq{ \lb{eq:55}
 \begin{split}
 &E = \frac{p^{2}_{z}}{2m} + \hbar\omega\big( 2r \big), \qquad r=0,1,2,3,\dots\\
 &\hspace{3cm}  n + m_{l} + 2m_{s} + 1 = 2r,
 \end{split}
}
where $r$ is associated to every energy level and known as the \emph{Landau level}.
Also, we can see easily that eigenstates with the same spatial probability density belong to different energy levels.
For instance, $F_{1,1}$ and $F_{1,-1}$ are associated to different energies, but their spatial probability density are equals.
In general we have,
\eq{
 \big|F_{n,m_{l}} \big|^{2} \!=\! \big|F_{n,-m_{l}} \big|^{2}, \quad E_{n,m_{l}}\!\neq\! E_{n,-m_{l}}.
}
In relation to Eq. (\ref{eq:55}), we can see that every Landau Level is highly degenerate, because the electron eigenstates with different quantum numbers $n$ and $m_{l}$ are related to the same energy level, as we see in Figure \ref{fig:niveis_oscilador_2D}.
This feature can be observed in the quantization of the cyclotron orbits of charged particles in magnetic field \cite{Mikhailov}.
Such charged particles can only occupy orbits with discrete energy values, but these levels are degenerate, where the degeneracy is associated to the number of electrons per level, which is directly proportional to the strength of the applied magnetic field.

Moreover, this physical aspect can be experimentally appreciated within solid materials, where there are electronic oscillations under the action of an external magnetic field; when we apply a differential of electric potential through the material, that leads to discrete values of electric current directly related to the Landau levels of the electrons.
This phenomenon is commonly called Integer Quantum Hall Effect \cite{Jeanneret}, and its useful applications have been evidenced in quantum metrology in order to acquire more information about microscopic details of semiconductors.
Also, evidence of Landau levels has been obtained in the propagation of electron vortex beams along an external longitudinal magnetic field \cite{Franco}.
\begin{figure}[t!]
 \centering
  \includegraphics[scale=0.17]{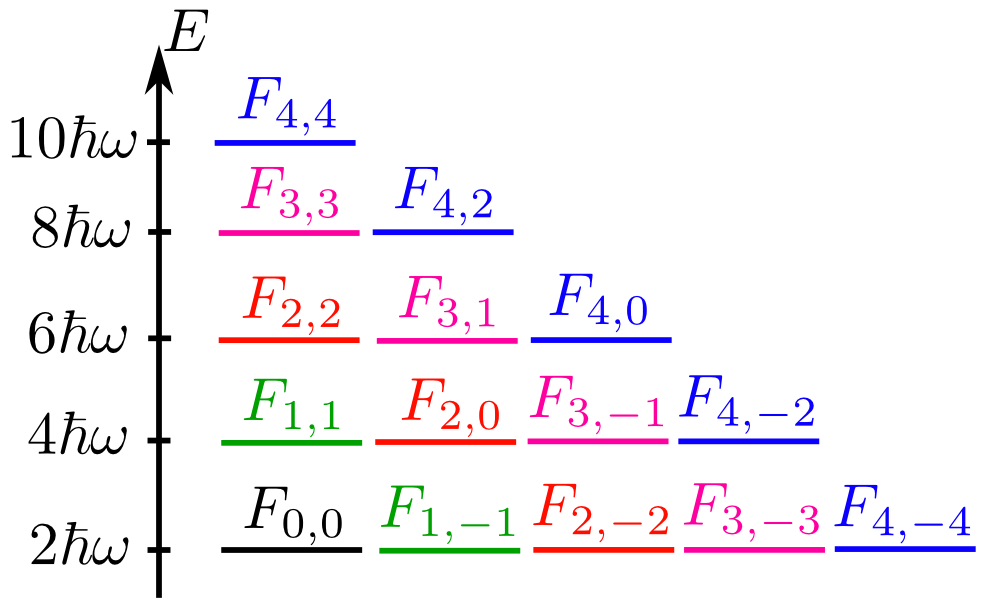}
  \caption{\small According to the first five values of $n$, its individual set of $m_{l}$ and $m_{s}\!=\!1/2$. We can see the degeneracy of the Landau levels.}
  \lb{fig:niveis_oscilador_2D}
\end{figure}
%
%
%
\section{Dirac equation for an electron in magnetic field} \lb{sec:IV}
In this section we present the relativistic description via Dirac equation for an electron in a region with a uniform magnetic field $\vec{B}$. According to this quantum description, the electron eigenstate is expressed by the spinorial formalism. 
In general, a Dirac spinor for the electron is a column vector with four-elements, each one of these are related to the eigenfunction (\ref{eq:51}) obtained from the relevant Schr\"odinger equation. 
Also, the order of the spinor's elements are associated with the spin orientation and the energy sign of the electron \cite{Strange,Greiner,Bransden}, where the energy negative elements are associated to the electron's antiparticle, commonly called positron.

After this short description, the electron Dirac equation is, 
\eq{\lb{eq:d-1}
 i\hbar\dfrac{\partial}{\partial t}U(\vec{r},t) \!=\! \big( c \tilde{\alpha}\cdot\hat{\vec{p}} + \tilde{\beta} m_{0}c^{2} \big)U(\vec{r},t), 
}
where $U$ is a general electron spinor.
We can write the linear momentum operator according to the canonical transformation  $\hat{\vec{p}} \!\rightarrow\! \hat{\vec{p}} + e\vec{A}$, where $\vec{A}$ is the magnetic vector potential.
Additionally, we have $\tilde{\alpha}$ and $\tilde{\beta}$ which are four Hermitian matrices with dimension $4\!\times\!4$, satisfying the condition $\tilde{\alpha}^{2}\!=\!\tilde{\beta}^{2}\!=\!\hat{\mathbb{1}}$,
\eq{
 \tilde{\alpha} \!=\! \left(\! \begin{array}{cc} 0 & \hat{\sigma}  \\  \hat{\sigma} & 0 \end{array}   \!\right), \quad \tilde{\beta} \!=\! \left(\! \begin{array}{cc} \hat{\mathbb{1}}_{2\times2} & 0 \\ 0 & -\hat{\mathbb{1}}_{2\times2} \end{array}  \!\right).  
}
Therefore, we obtain the time-independent Dirac Hamiltonian $\hat{H}\!=\!c \tilde{\alpha}\cdot\hat{\vec{p}} + \tilde{\beta} m_{0}c^{2}$.
We can express the electron's spinor evolution through an unitary transformation, 
\eq{\lb{eq:d-3}
 U(\vec{r},t) = \left( \! \begin{array}{c} \phi \\ \chi \end{array} \! \right)  e^{\resizebox{0.07\hsize}{!}{$\dst-i\frac{Et}{\hbar}$} },
}
where $\phi$ and $\chi$ are two scalar functions that represent the first two elements of positive energy and the last two elements of negative energy of the electron's spinor respectively.
When we substitute the spinor (\ref{eq:d-3}) in Eq. (\ref{eq:d-1}), and using the matrix notation, we obtain,
\eq{\lb{eq:d-4}
 \left( \!\! \begin{array}{cc}
 m_{0}c^{2}\hat{\mathbb{1}}_{2\times2} & \!\!c\hat{\sigma}\!\cdot\!(-i\hbar\nabla\!+\!e\vec{A}) \\
 c\hat{\sigma}\!\cdot\!(-i\hbar\nabla\!+\!e\vec{A})  &  -m_{0}c^{2}\hat{\mathbb{1}}_{2\times2}
 \end{array} \!\! \right) \!\!\left( \!\!\begin{array}{c} \phi \\ \chi \end{array} \!\! \right) \!=\! E \left(\!\!\begin{array}{c} \phi \\ \chi \end{array}\!\!\right).
}
From the first and second row of the matrix we can obtain two linear equations equal to the electron total energy times the respective scalar functions defined before,
\begin{subequations}\lb{eq:d-5}
 \begin{align}
	&m_{0}c^{2}\phi + c\hat{\sigma}\!\cdot\!(-i\hbar\nabla+e\vec{A})\chi = E\phi \lb{eq:d-5a}\\
	&c\hat{\sigma}\!\cdot\!(-i\hbar\nabla+e\vec{A})\phi - m_{0}c^{2}\chi = E\chi. \lb{eq:d-5b}
 \end{align}
\end{subequations}
When we isolate $\phi$ from (\ref{eq:d-5a}) and $\chi$ from (\ref{eq:d-5b}) we find both,
\eq{\lb{eq:d-6}
\phi = \frac{c\hat{\sigma}\!\cdot\!(-i\hbar\nabla+e\vec{A})}{E - m_{0}c^{2}}\chi, \quad \chi = \frac{c\hat{\sigma}\!\cdot\!(-i\hbar\nabla+e\vec{A})}{E + m_{0}c^{2}}\phi.
}
Replacing $\chi$ by $\phi$ and vice versa, results in,
\eq{\lb{eq:d-7}
\begin{split}
\hspace{-0.2cm} \frac{\big(\!E^{2} \!-\! m^{2}_{0}c^{4}\!\big)}{c^{2}}\phi &\!=\! \big[\hat{\sigma}\!\cdot\!(-\!i\hbar\nabla\!+\!e\vec{A})\big] \! \big[\hat{\sigma}\!\cdot\!(-\!i\hbar\nabla\!+\!e\vec{A})\big]\phi,\\
	&\!=\! \big[  (-\!i\hbar\nabla\!+\!e\vec{A})\!\cdot\!(-\!i\hbar\nabla\!+\!e\vec{A})\big]\phi \\
		&\;\;+\! i\hat{\sigma}\!\cdot\!\big[ (-\!i\hbar\nabla\!+\!e\vec{A}) \!\times\! (-\!i\hbar\nabla\!+\!e\vec{A})   \big]\phi.
\end{split}
}
The first term on the right is equal to
\eq{\lb{eq:d-8}
 \begin{split}
 \big[ \! (-\!i\hbar\nabla\!+\!e\vec{A}) \!\cdot\! (-\!i\hbar\nabla+e\vec{A}) \! \big]\phi &\!=\! -\hbar^{2}\nabla^{2}\phi + e^{2}\!A^{2}\phi \\
	&- ie\hbar\big( \nabla\!\cdot\!\vec{A} \!+\! \vec{A}\!\cdot\!\nabla \big)\phi.
 \end{split}
}
Similarly, the second term on the right side of Eq. (\ref{eq:d-7}) is equal to
\eq{\lb{eq:d-9}
 \begin{split}
 i\hat{\sigma}\!\cdot\!\big[\! (-i\hbar\nabla+e\vec{A}) \!\times\! (-i\hbar\nabla+e\vec{A})  \!\big]\phi &\!=\! e\hbar \hat{\sigma}\!\cdot\! \big[ \!(\nabla\!\times\!\vec{A}) \\ 
	&\;\;\;+\! (\vec{A}\!\times\!\nabla) \!\big]\phi.
 \end{split}
}
Then we obtain,
\eq{\lb{eq:d-10}
\begin{split}
 \big( \nabla\!\cdot\!\vec{A} \!+\! \vec{A}\!\cdot\!\nabla \big)\phi &\!=\! \cancel{ (\nabla\!\cdot\!\vec{A}) }\phi \!+\! \vec{A}\!\cdot\!\nabla\phi \!+\! \vec{A}\!\cdot\!\nabla\phi, \\
		&= \big[2\vec{A}\!\cdot\!\nabla\big]\phi,  \\
 \big[\! (\!\nabla\!\times\!\vec{A}\!) \!+\! (\!\vec{A}\!\times\!\nabla\!)\! \big]\phi &\!=\! (\!\nabla\!\times\!\vec{A}\!)\phi \!-\! \vec{A}\!\times\!(\!\nabla\phi\!) \!+\! \vec{A}\!\times\!(\!\nabla\phi\!), \\ 
		&= (\nabla\!\times\!\vec{A})\phi.
\end{split}
}
According to the last results, we rewrite Eq. (\ref{eq:d-7}), getting, 
\eq{\lb{eq:d-11}
 \begin{split}
 \frac{\big( E^{2} \!-\! m^{2}_{0}c^{4}\big)}{c^{2}}\phi &\!=\! \big[\!-\!\hbar^{2}\nabla^{2} + e^{2}\!A^{2} - 2i\hbar e(\vec{A}\!\cdot\!\nabla) \\
		&\qquad\qquad + e\hbar\hat{\sigma}\cdot\vec{B} \big]\phi.
 \end{split}
}
It's worth mentioning that $e\hbar\hat{\sigma}\cdot\vec{B}$ is the term associated with the interaction potential between the spin of the electron and the external magnetic field.
Usually, this term is put in by hand as we, see Eq. (\ref{eq:13}).
This is done in order to have a good quantum description by Schr\"odinger equation of an electron with spin $1/2$ in a region with an external magnetic field.
On the other hand, we have that this interaction potential appears naturally by using the Dirac equation to describe the same physical system.

Continuing the analysis, we make a simplification (without loss of physical generality), assuming an external uniform magnetic field along $\hat{z}$ axis.
In this way, we can express the magnetic vector potential by the Landau gauge which lets us to put in evidence the axial symmetry as we explained in last section,
\eq{\lb{eq:d-12}
 \vec{A} = \frac{1}{2}\big(-\!yB,xB,0\big).
}
Replacing the vector potential in Eq. (\ref{eq:d-11}) and dividing this equation by $2m_{0}$, we have, 
\eq{\lb{eq:d-14}
 \begin{split}
 \!\!\!\!\!\!&\left(\!\frac{ E^{2} \!-\! m^{2}_{0}c^{4}}{2m_{0}c^{2}} \!\right)\!\phi\\ 
 & \hspace{1.5cm}\!=\! \Bigg[\!\frac{-\hbar^{2}}{2m_{0}} \!\left( \!\frac{\partial^{2}}{\partial x^{2}} \!+\! \frac{\partial^{2}}{\partial y^{2}} \!\right) \!+\! \frac{e^{2}B^{2}}{8m_{0}}  \!\big( x^{2}\!+\!y^{2} \big)   \\
		&\hspace{2.5cm}   +\! \frac{eB}{2m_{0}}\! \big( \hat{L}_{z} \!+\! 2\hat{S}_{z} \big) \!-\! \frac{\hbar^{2}}{2m_{0}}\frac{\partial^{2}}{\partial z^{2}}  \! \Bigg]\phi.
 \end{split}
}
The first three terms in the right hand side correspond to the Hamiltonian of the two-dimensional harmonic oscillator and the magnetic field interaction between angular momentum and spin.
The fourth term corresponds to the Hamiltonian of a free particle along $\z$ axis.
We can express the first two elements of the electron spinor by the product of the eigenfunctions of the two-dimensional harmonic oscillator obtained in previous section, and a plane wave function of a free particle along the direction of the external field,
\eq{\lb{eq:d-15}
 \phi(\vec{r}) = F(x,y)e^{ip_{z}z/\hbar}\Gamma,
}
where $\Gamma$ is the spinorial function that we defined previously in Eq. (\ref{eq:spinorial_function}).
In relation to the first two spinor elements, which are related to the positive energy particle, we have the spin orientation up or down.
We further have the last two spinor elements which are related with negative energy particle (i.e., an antiparticle); these are related to the spin orientation as before.
Therefore, we can denote the two positive spinor elements as,
\eq{\lb{eq:d-16}
 \hspace{-0.1cm} U^{(+)}_{+\frac{1}{2}}(x,y) \!=\! \left(\!\! \begin{array}{c} F(x,y) \\ 0 \end{array} \!\!\right), \quad U^{(+)}_{-\frac{1}{2}}(x,y) \!=\! \left(\!\! \begin{array}{c} 0 \\ F(x,y)  \end{array} \!\!\right).
}
When the spin operator acts on the spinor positive energy elements, we obtain the relation,
\eq{\lb{eq:d-18}
 \hat{S}_{z}U^{(+)}_{m_{s}}(x,y) = \hbar m_{s}U^{(+)}_{m_{s}}(x,y), \quad  m_{s}=\left\{ \!\! \begin{array}{l} +1/2  \\ -1/2 \end{array}\right..
}
Additionally, the commutation relations between the axial angular momentum and the harmonic Hamiltonian are equal to zero, we can say that $\hat{L}_{z}$ is a conserved physical quantity, and its eigenfunctions are the eigenfunctions of the two-dimensional harmonic oscillator.
Thus, we get the following relation,
\eq{\lb{eq:d-19}
 \hat{L}_{z}U^{(+)}_{m_{s}}(x,y) = \hbar m_{l}U^{(+)}_{m_{s}}(x,y),
}
where $m_{l}$ is the quantum number of the axial angular momentum.
When we replace Eq. (\ref{eq:d-15}) in Eq. (\ref{eq:d-14}) we obtain,  
\eq{\lb{eq:d-20}
 \begin{split}
 \left(\!  \frac{E^{2} \!-\! m^{2}_{0}c^{4}}{2m_{0}c^{2}} \!\right)\!F &\!=\! \Bigg[ \frac{-\!\hbar^{2}}{2m_{0}} \!\left(\!\frac{\partial^{2}}{\partial x^{2}} \!+\! \frac{\partial^{2}}{\partial x^{2}}\!\right)\\
	& \quad +\! \frac{1}{2}m_{0}\omega^{2}\big(x^{2}\!+\!y^{2}\big) \!+\! \frac{p^{2}_{z}}{2m_{0}} \\
	& \hspace{1.8cm} +\! \hbar\omega\big(m_{l} \!+\! 2m_{s}\big) \Bigg]\!F.
 \end{split}
}
Rearranging the last equation,
\eq{\lb{eq:d-21}
 \begin{split}
	&\Bigg[\! \frac{-\hbar^{2}}{2m_{0}}\!\left(\frac{\partial^{2}}{\partial x^{2}}\!+\!\frac{\partial^{2}}{\partial x^{2}}\right) \!+ \frac{1}{2}m_{0}\omega^{2}\big(x^{2}\!+\!y^{2}\big) \!\Bigg]\!F \\
	&\qquad =\! \left[\! \frac{E^{2} \!-\! m^{2}_{0}c^{4}}{2m_{0}c^{2}} \!-\!\frac{p^{2}_{z}}{2m_{0}} \!-\! \hbar\omega\big(m_{l} \!+\! 2m_{s}\big) \!\right]\!F,
 \end{split}
}
we notice that the left hand terms in brackets correspond to the Hamiltonian of the two-dimensional harmonic oscillator in Cartesian coordinates with a characteristic frequency $\omega\!=\!eB/2m_{0}$.
Therefore, we can say that the right hand terms is equal to the energy of the oscillator.
We accordingly have,
\eq{\lb{eq:d-22} 
 \!\!\!\! \left[ \! \frac{E^{2} \!-\! m^{2}_{0}c^{4}}{2m_{0}c^{2}} \!-\! \frac{p^{2}_{z}}{2m_{0}} \!-\! \hbar\omega\big(m_{l} \!+\! 2m_{s}\big) \! \right] \!=\! \hbar\omega\big( \! n_{x} \!+\! n_{y} \!+\!1 \!\big).
}
Due to the total energy is to second power, when we isolate $E$, we obtain two signs for the total energy of the particle, being the positive sign related to the electron and the negative sign to the positron,
\eq{\lb{eq:d-23}
 E \!=\! \pm\sqrt{ m^{2}_{0}c^{4} \!+\! p^{2}_{z}c^{2} \!+\! eB\hbar c^{2}\big( n \!+\! m_{l} \!+\! 2m_{s} \!+\! 1 \big) }.   
}
If we have the particular case of an electron with null linear momentum, an angular momentum projection equal to $m_{l}\!=\!-n$ and a weak magnetic field in relation to the rest mass of the electron, then we can approximate the energy expression as,
\eq{
 \begin{split}
 E &\approx m_{0}c^{2} \left( 1 + \frac{eB\hbar}{2m^{2}_{0}c^{2}} \left(2m_{s} \!+\! 1\right) \right),\\
	& = m_{0}c^{2} + \frac{eB\hbar}{2m_{0}} \left(2m_{s} \!+\! 1\right).
 \end{split}
}
In relation to this result we obtain the energy levels of the spin interaction with an external magnetic field as we shown in the last two sections, also the energy shift related to the zero-point energy of the harmonic potential, and the rest mass of the electron, which is relevant by a relativistic description as we done using Dirac equation.

According to the axial symmetry exhibited by the two dimensional harmonic Hamiltonian Eq. (\ref{eq:21}), we can make a coordinate system transformation in order to rewrite the Hamiltonian in cylindrical coordinates, as we have done in previous section.
We thus obtain the same radial eigenfunctions of the electron, Eq. (\ref{eq:51}) in the case where the electron state is expressed by the Dirac spinor.

Specifically, if we describe the electron using the Dirac equation, we have to consider the first two elements of the Dirac spinor as the radial eigenfunctions obtained in last section.
Thus, we implement the relation (\ref{eq:d-6}), according to the orientation of the electron's spin.
In the same way, we can obtain the other two elements of the spinor which are related to negative energy.
Then, we express the complete Dirac spinor associated with a particular quantum state of the electron.

To exemplify this process, we calculate a particular case to obtain the Dirac spinor for an electron in a specific state.
We choose the ground state of the electron $n\!=\!0$, $m_{l}\!=\!0$ and its spin orientation $m_{s}\!=\!+1/2$.
Then, the electron relativistic energy is,  
\eq{
 E_{0,0} \!=\! \sqrt{ m^{2}_{0}c^{4} \!+\! p^{2}_{z}c^{2} \!+\! 2eB\hbar c^{2} }.  
}
\begin{figure}[t!]
 \centering
  \includegraphics[scale=0.25]{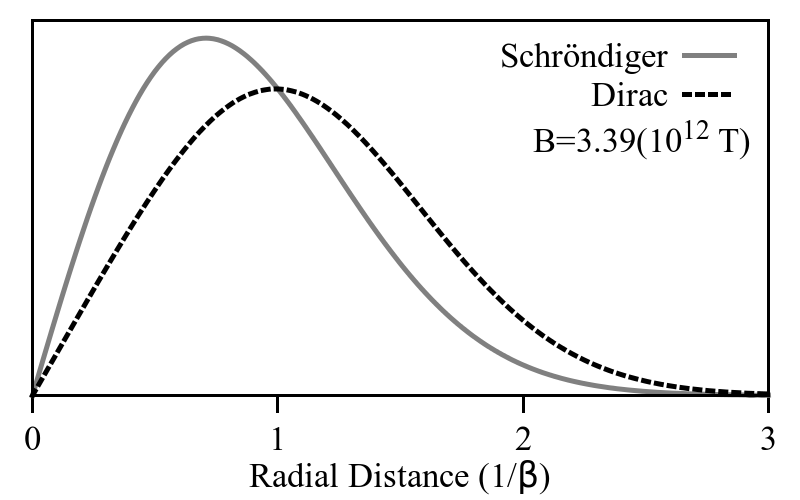}
 \caption{\small Comparison of radial probability density via Schr\"odinger and Dirac equation of the electron ground state.}
 \lb{fig:estado_base}
\end{figure}
\\
Previously, we have obtained the radial wave function of the ground state via Schr\"odinger equation,
\eq{
 F_{0,0}(\varrho,\varphi) = \frac{\beta}{\sqrt{\pi}}e^{-\beta^{2}\varrho^{2}/2}, \qquad \beta=\sqrt{\frac{eB}{2\hbar}}.
}
We can write the first two elements of the spinor,
\eq{
 U^{(+)}_{0,0}(\vec{r}) = \left( \! \begin{array}{c}F_{0,0}(\varrho,\varphi) \\ 0 \end{array} \! \right)e^{ip_{z}z/\hbar}.
}
when we transform the relations (\ref{eq:d-6}) in cylindrical coordinates.

Firstly, we have to express every unitary vectors of Cartesian coordinate as,
\eq{
 \left(\begin{array}{c} \x \\ \y \\ \z \end{array}\right) = \left(\begin{array}{ccc}  
 \cos\varphi & -\sin\varphi & 0\\
 \sin\varphi & \cos\varphi  & 0\\
     0       &      0       & 1 
\end{array} \right)  
\left(\begin{array}{c} \hat{\varrho} \\ \hat{\varphi} \\ \z \end{array}\right).
}
In this way, the magnetic vector potential is, 
\eq{
 \vec{A} =\frac{B\varrho}{2}\hat{\varphi}.
}
Similarly, the nabla operator in cylindrical coordinates is,
\eq{
 \nabla = \hat{\varrho}\frac{\partial}{\partial\varrho}+\frac{\hat{\varphi}}{\varrho}\frac{\partial}{\partial\varphi}+\hat{z}\frac{\partial}{\partial z}.
} 
Also, the Pauli matrices vector is written as,
\eq{
\begin{split}
 \vec{\sigma} &= \hat{\sigma}_{x}\x + \hat{\sigma}_{y}\y + \hat{\sigma}_{z}\z,\\
	&=\! \left( \!\! \begin{array}{cc} 0 & e^{-i\varphi}\\e^{i\varphi} & 0 \end{array} \!\! \right)\!\hat{\varrho} \!+\! \left( \!\! \begin{array}{cc} 0 & -ie^{-i\varphi}\\ie^{i\varphi} & 0 \end{array} \!\! \right)\!\hat{\varphi} \!+\! \left( \!\! \begin{array}{cc} 1 & 0\\0 & -1 \end{array} \!\! \right)\!\z.
\end{split}
}
Once we have calculated all the relevant transformations, we are able to use Eq.  (\ref{eq:d-6}) to calculate the two negative energy elements of the spinor as follows, 
\eq{
 \frac{\dst c\hat{\sigma}\!\cdot\!\big( \!-\!i\hbar\nabla \!+\! \frac{eB\varrho}{2}\hat{\varphi} \big) }{E\!+\!m_{0}c^{2}} \!=\! \frac{1}{E\!+\!m_{0}c^{2}} \!\!\! \left( \!\! \begin{array}{cc} \dst-i\hbar c\frac{\partial}{\partial z} & \!\!\dst -2i\hbar c\beta\hat{a}_{R}\\ \dst  2i\hbar c\beta\hat{a}^{\dagger}_{R} & \dst i\hbar c\frac{\partial}{\partial z} \end{array} \!\! \right),
}
where $\hat{a}^{\dagger}_{R}$ and $\hat{a}_{R}$ are the ladder operators defined in Eq. (\ref{eq:46}).
Therefore, the two electron spinor elements corresponding to the negative energy of the ground state can be expressed as,
\eq{\lb{eq:rc2_16}
  U^{(-)}_{0,0} = \frac{1}{E\!+\!m_{0}c^{2}} \left( \! \begin{array}{c} cp_{z}F_{0,0}(\varrho,\varphi) \\ 2i\hbar c\beta F_{1,1}(\varrho,\varphi) \end{array} \! \right) \! e^{ip_{z}z/\hbar}.
}
Therefore, we obtain the complete Dirac spinor associated to the ground state of the electron,
\eq{\lb{eq:rc2_17}
	U_{0,0}(\vec{r}) = N_{0,0} \left(\begin{array}{c}  F_{0,0}(\varrho,\varphi) \\ 0 \\ \dst\frac{cp_{z}F_{0,0}(\varrho,\varphi)}{(E_{0,0}\!+\!m_{0}c^{2})} \\ \dst\frac{2i\hbar c \beta F_{1,1}(\varrho,\varphi)}{(E_{0,0} \!+\! m_{0}c^{2})} \end{array}\right)e^{ip_{z}z/\hbar}, 
}
where its normalization constant is, 
\eq{\lb{eq:rc2_18}
 N_{0,0} = \left[ 1 \!+\! \frac{c^{2}p^{2}_{z}}{(E_{0,0}\!+\!m_{0}c^{2})^{2}} \!+\! \frac{4\hbar^{2} c^{2}\beta^{2}}{(E_{0,0}\!+\!m_{0}c^{2})^{2}} \right]^{-\frac{1}{2}}.
}
We would like to emphasize the relevance of the two negative elements of the Dirac spinor for an electron in two specific cases: \emph{i}) When the electron has a great linear momentum and it is the dominant physical quantity, we have that the electron speed is very close to the speed of light and the third spinor element becomes relevant for this description. \emph{ii}) When we have a strong confinement of the electron and it is trapped in a region less than or equal to its Compton wavelength \cite{Leary,Velasco}.
Thereby, the fourth spinor element becomes relevant to the physical description.
Thus, we can obtain this last case when the electron is in a region with a very strong magnetic field, yielding a high characteristic frequency $\omega$ of the harmonic potential, as consequence a very small natural length of the harmonic oscillator $\sqrt{\hbar/m_{0}\omega}$.
As such, we show in Figure \ref{fig:estado_base} a different radial probability density of the electron ground state via Dirac equation, in comparison to that obtained before via Schr\"odinger equation.

The fourth element of the spinor shows an excited eigenstate $n,m_{l}\!=\!1,1$; whose contribution is comparable to the first element.
Subsequently, this element of the spinor becomes relevant for the description of the electron quantum state; and this is why we observe a more open electron radial probability density via the Dirac equation.

We advise to the reader to see a short animation \cite{animation} about the radial probability density according to both theories in relation to the external magnetic field strength.

In summary, we can conclude that the quantum mechanical description via the Dirac equation requires more complicated mathematical steps, but that it also brings new concepts, such as the negative energy of the electron and a different treatment through the application of Dirac spinors; this yields a more general way to describe the quantum state of an electron when relativistic effects are taken into account which is richer than the Schr\"odinger description.

\section{Acknowledgements} 
The first author wish to thank to his doctoral advisor Eduardo, for all fruitful discussions and his great listening and grateful understanding. Also to Benjamin for his great collaboration and finally the Brazilian agency CAPES for its financial support.



\begin{thebibliography}{99}

   \bibitem{Peter_Rowe}{E. G. Peter Rowe, \emph{Classical limit of quantum mechanics (electron in a magnetic field)}, \href{http://scitation.aip.org/content/aapt/journal/ajp/59/12/10.1119/1.16622}{Am. J. Phys. \textbf{59}, 1111 (1991)}.}
   \bibitem{Tannoudji}{C. Cohen-Tannoudji, B. Diu, F. Laloë; \emph{Quantum mechanics} vol.1 Section D$_{\tn{VI}}$, \href{http://www.wiley.com/WileyCDA/WileyTitle/productCd-047116433X.html}{\textbf{ISBN}: 978-0-471-16433-3, Wiley (1991).}}
   \bibitem{Haugset}{T. Haugset, J. Aa. Ruud and F. Ravndal; \emph{Gauge invariance of Landau levels}, \href{http://iopscience.iop.org/article/10.1088/0031-8949/47/6/004}{Physica Scripta \textbf{47}, 715 (1993).}}
   \bibitem{Yoshioka}{D. Yoshioka, \emph{The Quantum Hall Effect}, Chapter 2, \href{http://www.springer.com/us/book/9783540431152}{\textbf{ISBN}: 978-3-642-07720-3, Springer (2002).}}
   \bibitem{Thaller}{B. Thaller; \emph{Advanced Visual Quantum Mechanics}, \href{http://www.springer.com/us/book/9780387207773}{\textbf{ISBN}: 0-387-20777-5, Springer (2004).} }
   \bibitem{Greenshields}{C. R. Greenshields, R. L. Stamps, S. Franke-Arnold, and S. M. Barnett, \emph{Is the Angular Momentum of an Electron Conserved in a Uniform Magnetic Field?}, \href{http://journals.aps.org/prl/abstract/10.1103/PhysRevLett.113.240404}{Phys. Rev. Lett. \textbf{113}, 240404 (2014).}}




   \bibitem{Babiker}{M. Babiker, J. Yuan, and V. E. Lembessis, \emph{Electron vortex beams subject to static magnetic fields}, \href{http://journals.aps.org/pra/abstract/10.1103/PhysRevA.91.013806}{Phys. Rev. A \textbf{91}, 013806 (2015).}}
   \bibitem{Chowdhury}{D. Chowdhury, B. Basu, and P. Bandyopadhyay, \emph{Electron vortex beams in a magnetic field and spin filter}, \href{http://journals.aps.org/pra/abstract/10.1103/PhysRevA.91.033812}{Phys. Rev. A \textbf{91}, 033812 (2015).}}
   \bibitem{Bliokh}{K. Bliokh et al., \emph{Semiclassical Dynamics of Electron Wave Packet States with Phase Vortices}, \href{http://journals.aps.org/prl/abstract/10.1103/PhysRevLett.99.190404}{Phys. Rev. Lett. \textbf{99}, 190404 (2007)}}




   \bibitem{Landau}{L. Landau, \emph{Diamagnetismus der Metalle}, \href{http://link.springer.com/article/10.1007%2FBF01397213}{Z. Physik \textbf{64}, 629-637 (1930).}}
   \bibitem{Feldman_Kahn}{A. Feldman and A. Kahn, \emph{Landau Diamagnetism from the Coherent State of an Electron in a Uniform Magnetic Field}, \href{http://journals.aps.org/prb/abstract/10.1103/PhysRevB.1.4584}{Phys. Rev. B \textbf{1}, 4584 (1970).}}
   \bibitem{Pekka}{T. Chakraborty, P. Pietiäinen; \emph{The Quantum Hall Effect: Integral and Fractional}, \href{http://download.springer.com/static/pdf/45/bfm%253A978-3-642-79319-6%252F1.pdf?originUrl=http%3A%2F%2Flink.springer.com%2Fbook%2Fbfm%3A978-3-642-79319-6%2F1&token2=exp=1463159409~acl=%2Fstatic%2Fpdf%2F45%2Fbfm%25253A978-3-642-79319-6%25252F1.pdf%3ForiginUrl%3Dhttp%253A%252F%252Flink.springer.com%252Fbook%252Fbfm%253A978-3-642-79319-6%252F1*~hmac=67fb60d6546548052d3d37e1c6d33e69668475eff316ba0c6e74f7026ce0f3d9}{\textbf{ISBN}: 13:978-3-540-58515-2, Springer (1995).} }
   \bibitem{laughlin}{R. B. Laughlin; \emph{Quantized Hall conductivity in two dimensions}, \href{http://journals.aps.org/prb/abstract/10.1103/PhysRevB.23.5632}{Phys. Rev. B \textbf{23}, 5632 (1981).}}
   \bibitem{Schattschneider}{P. Schattschneider et al, \emph{Imaging the dynamics of free-electron Landau states}, \href{http://www.nature.com/ncomms/2014/140808/ncomms5586/full/ncomms5586.html}{Nat. Commun. \textbf{5}, 4586 (2014).}}

   \bibitem{Ole}{Ole Steuernagel, \emph{Equivalence between focused paraxial beams and the quantum harmonic oscillator},  \href{http://scitation.aip.org/content/aapt/journal/ajp/73/7/10.1119/1.1900099}{Am. J. Phys. \textbf{73}, 625 (2005).}}

   \bibitem{Allen}{L. Allen et al., \emph{Optical Angular Momentum of light and the transformation of Laguerre-Gaussian laser modes}, \href{http://journals.aps.org/pra/abstract/10.1103/PhysRevA.45.8185}{Phys. Rev. A \textbf{45}, 8185 (1992).}}
   \bibitem{Zhang}{S. Zhang and Z. Yang, \emph{Intrinsic Spin and Orbital Angular Momentum Hall Effect}, \href{http://journals.aps.org/prl/abstract/10.1103/PhysRevLett.94.066602}{Phys. Rev. Lett. \textbf{94}, 066602 (2005).}}


   \bibitem{Jones_vector}{W. E. Baylis, J. Bonenfant, J. Derbyshire and J. Huschilt, \emph{Light polarization: A geometric-algebra approach}, \href{http://scitation.aip.org/content/aapt/journal/ajp/61/6/10.1119/1.17205}{Am. J. Phys. \textbf{61}, 534 (1993).}}


   \bibitem{Goubau}{G. Goubau and F. Schwering, \emph{On the guided propagation of electromagnetic wave beams}, \href{http://ieeexplore.ieee.org/xpls/abs_all.jsp?arnumber=1144999&tag=1}{IRE Transactions on Antennas and Propagation \textbf{9}, 248 (1961).}}
   \bibitem{Andrews}{David L. Andrews and Mohamed Babiker; \emph{The Angular Momentum of Light}, 1st edition, \href{https://books.google.com.br/books?id=li2bysEqHf0C&pg=PR4&lpg=PR4&dq=ISBN+978-1-107-00634-8&source=bl&ots=JPHba07_MJ&sig=Bk2dzsRrvMez17N1RkfakkRLLOM&hl=pt-BR&sa=X&ved=0ahUKEwjf_sLtw9fMAhXCCpAKHTuJCRcQ6AEIHDAA#v=onepage&q=ISBN%20978-1-107-00634-8&f=false}{\textbf{ISBN}:  978-1-107-00634-8, Cambridge University Press (2013).}}
   \bibitem{Simin_Feng}{S. Feng and H. G. Winful, \emph{Physical origin of the Gouy phase shift}, \href{https://www.researchgate.net/profile/Simin_Feng2/publication/5808178_Physical_Origin_of_the_Gouy_Phase_Shift/links/53d9d41b0cf2631430c7dd1a.pdf}{Opt. Lett. \textbf{26}, 485-487 (2001).}}
   \bibitem{Paula}{Paula B. Monteiro, Paulo A. Maia Neto, and H. Moys\'es Nussenzveig, \emph{Angular momentum of focused beams: Beyond the paraxial approximation}, \href{http://journals.aps.org/pra/abstract/10.1103/PhysRevA.79.033830}{Phys. Rev. A \textbf{79}, 033830 (2009).}}
   \bibitem{Mikhailov}{S.A. Mikhailov, \emph{A new approach to the ground state of quantum Hall systems. Basic principles}, \href{http://www.sciencedirect.com/science/article/pii/S0921452600007699}{Physica B \textbf{299}, 6-31 (2001).}}   

   \bibitem{Jeanneret}{B. Jeanneret, B. D. Hall, H.J. B\"uhlmann, R. Houdr\'e, M. Ilegems, B. Jeckelmann, and U. Feller; \emph{Observation of the integer quantum Hall effect by magnetic coupling to a Corbino ring}, \href{http://journals.aps.org/prb/abstract/10.1103/PhysRevB.51.9752}{Phys. Rev. B \textbf{51}, 9752 (1995).}}


   \bibitem{Franco}{K. Y. Bliokh, P. Schattschneider, J. Verbeeck, and F. Nori; \emph{Electron Vortex Beams in a Magnetic Field: A New Twist on Landau Levels and Aharonov-Bohm States}, \href{http://journals.aps.org/prx/abstract/10.1103/PhysRevX.2.041011}{Phys. Rev. X \textbf{2}, 041011 (2012).}}




   \bibitem{Strange}{P. Strange; \emph{Relativistic Quantum Mechanics with applications in condensed matter and atomic physics}, 1st edition, \href{http://www.directtextbook.com/isbn/9780521565837}{\textbf{ISBN}: 978-0-521-56583-7, Cambridge University Press (1998).}}
   \bibitem{Greiner}{W. Greiner; \emph{Relativistic Quantum Mechanics wave equations}, 3rd edition, \href{https://books.google.com.br/books?id=2DAInxwvlHYC&pg=PR4&lpg=PR4&dq=ISBN+3-540-67457-8&source=bl&ots=5t0tfbIYT8&sig=xsvb5uD_WUtsB9J3InrOgTC4lRs&hl=pt-BR&sa=X&ved=0ahUKEwjsl-Sux9fMAhWBIpAKHdQoASwQ6AEIHDAA#v=onepage&q=ISBN%203-540-67457-8&f=false}{\textbf{ISBN}: 3-540-67457-8, Springer (2000).}}
   \bibitem{Bransden}{B. H. Bransden and C. J. Joachain; \emph{Quantum Mechanics}, 2nd edition, \href{http://www.amazon.com/Quantum-Mechanics-2nd-B-H-Bransden/dp/0582356911}{\textbf{ISBN}: 978-0582356917, Prentice Hall (2000).}}





  

   \bibitem{Leary}{C. C. Leary and Karl H. Smith, \emph{Unified dynamics of electrons and photons via Zitterbewegung and spin-orbit interaction}, \href{http://journals.aps.org/pra/abstract/10.1103/PhysRevA.89.023831}{Phys. Rev. A \textbf{89}, 023831 (2014).}}
   \bibitem{Velasco}{D. V. Villamizar and E. I. Duzzioni, \emph{Quantum speed limit for a relativistic electron in a uniform magnetic field}, \href{http://journals.aps.org/pra/abstract/10.1103/PhysRevA.92.042106}{Phys. Rev. A \textbf{92}, 042106 (2015)}.}
   \bibitem{animation}{Find in YouTube: \emph{Comparison Schr\"odinger and Dirac Radial Probability Density} or \url{https://youtu.be/o0cvFce1y10}}


\end{thebibliography}
\end{document}